
\magnification\magstephalf

\def\k{{\bf k}}
\def\x{{\bf x}}
\def\A{{\bf A}}
\def\o{|\k |}
\font\rfont=cmr10 at 10 true pt
\def\ref#1{$^{\hbox{\rfont {[#1]}}}$}

\font\fourteenrm=cmr12 scaled\magstep1

   \def \d {\delta}
 
 \def\t {\theta} \def\la{\lambda}
  
\def\pd {\partial}
\def\pmb#1{\setbox0=\hbox{#1}
 \kern.05em\copy0\kern-\wd0 \kern-.025em\raise.0433em\box0 }

\def \half {{\scriptstyle {1 \over 2}}}

\def \quarter {{\scriptstyle {1 \over 4}}}

 %


\def\boxit#1{\vbox{\hrule\hbox{\vrule\kern1pt\vbox
{\kern1pt#1\kern1pt}\kern1pt\vrule}\hrule}}

\parskip=6pt
\parindent=0pt
\hsize=17truecm\hoffset=-5truemm
\voffset=-1truecm\vsize=26.5truecm
\def\footnoterule{\kern-3pt
\hrule width 17truecm \kern 2.6pt}


\catcode`\@=11 

\def\nolabels{\def\wrlabeL##1{}\def\eqlabeL##1{}\def\reflabeL##1{}}
\def\writelabels{\def\wrlabeL##1{\leavevmode\vadjust{\rlap{\smash%
{\line{{\escapechar=` \hfill\rlap{\sevenrm\hskip.03in\string##1}}}}}}}%
\def\eqlabeL##1{{\escapechar-1\rlap{\sevenrm\hskip.05in\string##1}}}%
\def\reflabeL##1{\noexpand\llap{\noexpand\sevenrm\string\string\string##1}}}
\nolabels
\global\newcount\refno \global\refno=1
\newwrite\rfile
\def\defref{$^{{\hbox{\rfont [\the\refno]}}}$\nref}
\def\nref#1{\xdef#1{\the\refno}\writedef{#1\leftbracket#1}%
\ifnum\refno=1\immediate\openout\rfile=refs.tmp\fi
\global\advance\refno by1\chardef\wfile=\rfile\immediate
\write\rfile{\noexpand\item{#1\ }\reflabeL{#1\hskip.31in}\pctsign}\findarg}
\def\findarg#1#{\begingroup\obeylines\newlinechar=`\^^M\pass@rg}
{\obeylines\gdef\pass@rg#1{\writ@line\relax #1^^M\hbox{}^^M}%
\gdef\writ@line#1^^M{\expandafter\toks0\expandafter{\striprel@x #1}%
\edef\next{\the\toks0}\ifx\next\em@rk\let\next=\endgroup\else\ifx\next\empty%
\else\immediate\write\wfile{\the\toks0}\fi\let\next=\writ@line\fi\next\relax}}
\def\striprel@x#1{} \def\em@rk{\hbox{}}
\def\lref{\begingroup\obeylines\lr@f}
\def\lr@f#1#2{\gdef#1{\defref#1{#2}}\endgroup\unskip}
\newwrite\lfile
{\escapechar-1\xdef\pctsign{\string\%}\xdef\leftbracket{\string\{}
\xdef\rightbracket{\string\}}}

\def\writestop{\def\writestoppt{\immediate\write\lfile{\string\p
ageno%
\the\pageno\string\startrefs\leftbracket\the\refno\rightbracket%
\string\def\string\secsym\leftbracket\secsym\rightbracket%
\string\secno\the\secno\string\meqno\the\meqno}\immediate\closeout\lfile}}
\def\writestoppt{}\def\writedef#1{}
\catcode`\@=12 
\rightline{DAMTP 93 -- 33}
\rightline{ITP -- SB-- 93 -- 38}
\bigskip
\centerline{\fourteenrm Canonical Quantisation in n.A=0 gauges}
\bigskip
\bigskip
\centerline{P V Landshof{}f}
\centerline{DAMTP, University of Cambridge}
\bigskip
\centerline{P van Nieuwenhuizen}
\centerline{Institute for Theoretical Physics, State University of New York at
Stony Brook}
\centerline{and}
\centerline{Teyler Foundation, Instituut Lorentz, Leiden}
\vskip 15truemm
{\bf Abstract}

We give a unified derivation of the propagator in the gauges $n.A=0$ for
$n^2$ timelike, spacelike or lightlike. We discuss the physical states
and other physical questions.

\vskip 15truemm
{\bf 1 Introduction}

Gauges of the type $n.A=0$ are widely used, with $n$ either timelike,
spacelike or lightlike. They often simplify calculations, and
give a  more direct physical  interpretation to Feynman
diagrams. An example is the derivation of the Altarelli-Parisi equation
for deep inelastic scattering\defref\rgr{
R G Roberts, {\sl The structure of the proton},Cambridge University
Press (1990)
}, where the ladder graphs without crossed rungs
dominate,  or finite-temperature field theory where
the heat bath already breaks the Lorentz invariance\defref\heinz{
U Heinz, K Kajantie and T Toimela, Ann Phys 176 (1987) 218
}.
It is useful also to use an axial gauge for the renormalisation of
composite operators, to avoid mixing with non-gauge-invariant operators (which
in general contain ghosts)\defref\zero{
W Furmanski and R Petronzio, Physics Letters B97 (1980) 437;
W van Neerven, Z Phys C14 (1982) 241
}. But it has been surprisingly
difficult to derive correct perturbation theory for these gauges\defref\one{
S Carracciolo, G Curci and P Menotti, Physics Letters 113B (1982) 311
}\defref\two{
P V Landshof{}f, review talk in  Proc Workshop on Nonstandard
Gauges, Vienna 1989, Springer Lecture Notes in Physics
}. This is because the Feynman propagator naively is\footnote{$^{\dag}$}{
Our metric is (+,--,--,--), so that $k^2=k_0^2-\o ^2$, $n^2=1-\lambda ^2$,
$\k .\x =-k_1x^1-k_2x^2-k_3x^3$, and $n.k=k_0+\lambda k_3$.
Furthermore, $\k .\x =k^jx^j$ and $\pd _j\pd _j=\pd _1^2+\pd _2^2+\pd _3^2$}
$$
D_{\mu\nu}(k)=\left [-g_{\mu\nu}+{{k_{\mu}n_{\nu}+n_{\mu}k_{\nu}}\over{n.k}}
-n^2{{k_{\mu}k_{\nu}}\over{(n.k)^2}}\right ]{1\over{k^2+i\epsilon}}
\eqno(1.1)
$$
and one must decide how to integrate the pole and double pole at $n.k=0$.
In this note we shall show that, in the gauges $A_0+\lambda A_3=0$, for all
$\lambda$ and so regardless of whether $n$ is timelike,
spacelike or lightlike, straightforward canonical quantisation
leads to
$$
{{\hbox{``}}\atop{~}}{1\over{n.k}}{{\hbox{''}}\atop{~}}
={1\over{k_0+\lambda k_3 \pm i\epsilon /k_3}}
\eqno(1.2)
$$
with $\epsilon$ an infinitesimally small positive quantity.

We derive this result in Section 2. In Section 3 we construct the physical
states and in Section 4 we discuss various physical questions.
\bigskip
{\bf 2 Derivation of the propagator}

To derive the propagator it is sufficient to consider the interaction-free
case.
That is, we work with the asymptotic $in$ or $out$ field, described by the
Lagrangian $-\quarter F_{\mu\nu}F^{\mu\nu}$, where
$F_{\mu\nu}=\pd _{\mu}A_{\nu}-\pd _{\nu}A_{\mu}$. Rather than adding a
gauge-fixing term, we fix the gauge by eliminating $A_0$ and replacing
it with $-\lambda A_3$. The resulting three field equations read
\goodbreak
$$
\ddot A_r+\lambda\pd _r\dot A_3-\pd _j\pd _jA_r+\pd _r(\pd _jA_j)=0
{}~~~~~~~~~~~(r=1,2;\; j=1,2,3)
$$$$
\ddot A_3+\lambda\pd _3\dot A_3+(\lambda ^2-1)\pd _j\pd _jA_3+\pd _3(\pd _jA_j)
+\lambda\pd _t(\pd _jA_j)=0
\eqno(2.1)
$$
These equations are easy to solve if one assumes that the solution may be
written as a four-dimensional Fourier integral, with $e^{-ik.x}=
e^{-ik_0t+i\k .\x}$. Eliminating $\k .\A (\k )$
gives
$$
(k_0+\lambda k_3)^2(k_0^2-\k ^2)A_3(k)=0
\eqno(2.2a)
$$
or
$$
(\pd _t+\lambda \pd _3)^2(\pd _t^2-\pd _j\pd _j)A_3(t,\x  )=0
\eqno(2.2b)
$$
(We always use $i$ and $j$ to range over the values 1,2,3, and $r$ and $s$
to range over 1,2.)
The general real solution of this equation may be written
$$
A_3(t,\x )=\int {{d^3k}\over{(2\pi )^3}}\left\{\t (k_3)
\left ( (itp_3(\k )+q_3(\k )\right )
e^{i(\la k_3t+\k .\x )}
+{1\over{2\o }}a_3(\k )e^{-i(\o t-\k .\x )}\right\}\;\;
+{\hbox{h c}}
\eqno(2.3)
$$
The frequencies $-\la k_3$ in the first term range from $-\infty$ to $+\infty$,
and so when we
add on the hermitian conjugate we have to include the $\theta(k _3)$ to
avoid double counting.

By making a similar decomposition of $A_r(t,\x )$, but with functions
$p_r (\k ),q_r (\k )$ and $a_r (\k )$, and substituting these expressions
for $A_3$ and $A_r$ back into the field equations (2.1), one finds
the general solution of the classical field equations
$$
A_r(t,\x )=\int {{d^3k}\over{(2\pi )^3}}\left\{\t (k_3)
k_r\left (it(\la ^2-1)p(\k )+q(\k )\right )e^{i(\la k_3t+\k .\x )}
{~~~~~~~~~~~}\right .
$$$$
{}~~~~~~~~~~~~~~~~~~~~~~~~~~~~~~~~~~~~~~~~~\left .
+{1\over{2\o }}a_r(\k )e^{-i(\o t-\k .\x )}\right\}
\;\;+{\hbox{h c}}
\eqno(2.4a)
$$$$
A_3(t,\x )=\int {{d^3k}\over{(2\pi )^3}}\left\{\t (k_3)
\left (\left (it(\la ^2-1)k_3+\la\right )p(\k )+k_3q(\k )\right )
e^{i(\la k_3t+\k .\x )}~~~~~~~~~~~~~~~~
\right .
$$$$
\left .
{}~~~~~~~~~~~~~~~~~~~~~~~~~~~~~~~~~~~~~~
-{1\over{2\o }}{{k_ra_r(\k )}\over{\la \o +k_3}}e^{-i( \o t-\k .\x
)}\right\}\;\;
+{\hbox{h c}}
\eqno(2.4b)
$$

To quantise, we construct the canonically-conjugate momenta
$\pi _i=\dot A_i-\lambda\pd _iA_3$:
$$
\pi _r(t,\x )=-i\int {{d^3k}\over{(2\pi )^3}}\left\{\t (k_3)
k_rp(\k )e^{i(\la k_3t+\k .\x )}~~~~~~~~~~~~~~~~~~~~~~~~
\right .
$$$$
\left .
{}~~~~~~~~~~~~~~~~~~~~~~~~~~~~~~+{1\over{2\o}}
\left (\o\d _{rs}-{{\la k_rk_s}\over{\la\o +k_3}}\right )a_s(\k )e^{-i(\o t-\k
.\x )}
\right\}\;\;+{\hbox{h c}}
\eqno(2.5a)
$$$$
\pi _3(t,\x )=i\int {{d^3k}\over{(2\pi )^3}}\left\{\t (k_3)
k_3p(\k )(\la ^2-1)e^{i(\la k_3t+\k .\x )}\right .
{}~~~~~~~~~~~~~~~~~~~~~~~~~~~~~~
$$$$
{}~~~~~~~~~~~~~~~~~~~~~~~~~~~~~~\left .+
{1\over{2\o}}\left ({{\o+\la k_3}\over{\la \o+k_3}}\right )k_ra_r(\k )
e^{-i(\o t-\k .\x )}\right\}\;\;+{\hbox{h c}}
\eqno(2.5b)
$$
With these solutions, the equal-time commutators involve terms with
$t^2$, $t$ and $\theta (k_3)$. In order to obtain  the canonical form,
the $t^2$ and $t$ terms should vanish, whereas the $\theta (k_3)$
should combine with $\theta (-k_3)$ to yield unity. The solution to all
these conditions is
\goodbreak
$$
[a_r(\k ),a_s^{\dag}(\k ')]=(2\pi )^3\d (\k -\k ')2\o\left\{
\d _{rs}+(\la ^2-1){{k_rk_s}\over{(\o +\la k_3)^2}}\right\}
$$$$
[p(\k ),p^{\dag}(\k ')]=0
$$$$
[q(\k ),q^{\dag}(\k ')]=(2\pi )^3\d (\k -\k ')\la (\la ^2-1)2k_3
\left ({1\over{\o ^2-\la ^2k_3^2}}\right )^2
$$$$
[q(\k ),p^{\dag}(\k ')]=(2\pi )^3\d (\k -\k '){1\over{\o ^2-\la ^2k_3^2}}
\eqno(2.6)
$$
Further, the $a_r$ sector commutes with the $p,q$ sector.

In order to derive the Feynman propagator, we must define the vacuum. The
issue of what vacua are allowed, and how to define the other physical states
in the gauges $n.A=0$,
is considered in detail in the next section. Meanwhile, we choose one of
the allowed vacua, defined by
$$
a_r(\k )|0\rangle =0;~~~~\;p(\k )|0\rangle =0;~~~~\;q(\k )|0\rangle =0
\eqno(2.7)
$$
Then, for $x^0>y^0$
\def\y{{\bf y}}
$$
\langle 0|TA_r(x^0,\x )A_s(y^0,\y )|0\rangle=\int {{d^3k}\over{(2\pi )^3}}
e^{i\k .(\x -\y )}\left [
\left \{\delta _{rs}+{{(\la ^2-1)k_rk_s}\over{(\o +\la k_3)^2}}\right \}
{1\over{2\o}}e^{-i\o (x^0-y^0)}
\right .~~~~~~~
$$$$
{}~~~~~~\left .
+e^{i\la k_3(x^0-y^0)}\theta (k_3)
\left \{ i(x^0-y^0)(\la ^2-1){{k_rk_s}\over{\o ^2-\la ^2k_3^2}}
+{{2k_rk_sk_3\la (\la ^2-1)}\over{(\o ^2-\la ^2k_3^2)^2}}
\right\}\right ]
\eqno(2.8a)
$$

$$
\langle 0|TA_r(x^0,\x )A_3(y^0,\y )|0\rangle=\int {{d^3k}\over{(2\pi )^3}}
e^{i\k .(\x -\y )}\left [
\left \{ {{-k_r(\la\o +k_3)}\over{(\o +\la k_3)^2}}\right \}
{1\over{2\o}}e^{-i\o (x^0-y^0)}
\right .~~~~~~~
$$$$
{}~~~~~~\left .
+e^{i\la k_3(x^0-y^0)}\theta (k_3)
\left \{ i(x^0-y^0)(\la ^2-1){{k_rk_3}\over{\o ^2-\la ^2k_3^2}}
+{{2k_rk_3^2\la (\la ^2-1)}\over{(\o ^2-\la ^2k_3^2)^2}}
+{{\la k_r}\over{\o ^2-\la ^2k_3^2}}
\right\}\right ]
\eqno(2.8b)
$$

$$
\langle 0|TA_3(x^0,\x )A_3(y^0,\y )|0\rangle=\int {{d^3k}\over{(2\pi )^3}}
e^{i\k .(\x -\y )}\left [
\left \{ {{k_sk_s}\over{(\o +\la k_3)^2}}\right \}
{1\over{2\o}}e^{-i\o (x^0-y^0)}
\right .~~~~~~~
$$$$
{}~~~~~~\left .
+e^{i\la k_3(x^0-y^0)}\theta (k_3)
\left \{ i(x^0-y^0)(\la ^2-1){{k_3^2}\over{\o ^2-\la ^2k_3^2}}
+{{2k_3^3\la (\la ^2-1)}\over{(\o ^2-\la ^2k_3^2)^2}}
+{{2\la k_3}\over{\o ^2-\la ^2k_3^2}}
\right\}\right ]
\eqno(2.8c)
$$
For $y^0>x^0$ we simply interchange the four-vectors $x$ and $y$.

For Feynman diagrams, we must convert the three-dimensional momentum integral
into a four-dimensional integral. Consider first the $(x^0-y^0)$ term
in (2.8a). If we make the replacement
$$
e^{i\la k_3(x^0-y^0)}\theta (k_3)
i(x^0-y^0)(\la ^2-1){{k_rk_s}\over{\o ^2-\la ^2k_3^2}}~~~~~~~~~~~~~~~~~~
{}~~~~~~~~~~~~~~~~~~~
$$$$
{}~~~~~~~~~~~~~~\longrightarrow{i\over{2\pi}}\int dk_0e^{-ik_0(x^0-y^0)}
{{(\la ^2-1)k_rk_s}\over{(k_0+\la k_3+i\epsilon/k_3)^2}}
{1\over{(k_0^2-\o ^2)+i\epsilon}}
\eqno(2.9)
$$
with $\epsilon$ and infinitesimal positive quantity,
the $i\epsilon/k_3$ correctly reproduces the $\theta (k_3)$ when we close
the contour of the $k_0$ integration. The residue of the double pole
reproduces the left-hand side of (2.9), together with another term
obtained by differentiating the last factor with respect to $k_0$.
This latter term agrees with the last term
in (2.8a). There are also the poles at $k_0=\pm \o$, whose
residue reproduces terms involving the other exponential in (2.8a).
We could proceed by combining them with a similar term
obtained by replacing the factor $1/2\o$ in the first term of (2.8a)
by a contour integral over $k_0$ involving again
$(k_0^2-\o ^2)+i\epsilon)^{-1}$.
However, at this point it is faster to note that the denominator
$(k_0+\la k_3+i\epsilon/k_3)^{-2}$ is just one of the two possible forms of
$``(n.k)$''$^{-2}$ as given
in (1.2). This suggests that all propagators in (2.8) are given by
four-dimensional Fourier transforms of (1.1) with this interpretation
of the denominators. This may be verified explicitly. To obtain
also the propagators involving $A_0$ we write it as $-\lambda A_3$. Hence
finally (for the vacuum defined in (2.7))
$$
\langle 0|TA_{\mu}(x^0,\x )A_{\nu}(y^0,\y )|0\rangle
=i\int {{d^4k}\over{(2\pi )^4}}e^{-ik.(x-y)}
\left [-g_{\mu\nu}+{{k_{\mu}n_{\nu}+n_{\mu}k_{\nu}}\over{n.k+i\epsilon/k_3}}
-n^2{{k_{\mu}k_{\nu}}\over{(n.k+i\epsilon/k_3)^2}}\right
]{1\over{k^2+i\epsilon}}
\eqno(2.10)
$$
For $\lambda <0$ Wick rotation is possible; for $\lambda >0$ see section 4.
\bigskip
{\bf 3 Definition of physical states}

A general approach to the definition of the physical states is in terms
of the BRST operator. When, as we have done, one of the fields has been
eliminated, an alternative is to use the lost field equation (here the Gauss
law) as a constraint that helps to pick out the physical states.

To derive the BRST operator we write the full Lagrangian in terms of
the Heisenberg fields, with a gauge-fixing term involving an auxiliary
field $B$ and with ghost fields:
$$
{\cal L}=-\quarter F_{a\mu\nu}F^{a\mu\nu}+B_an.A^a+b_an^{\mu}(D_{\mu}c)^a
\eqno(3.1)
$$
where
$(D_{\mu}c)^a=\pd _{\mu}c^a+gf^a_{bc}A^b_{\mu}c^c$. We take the auxiliary
field $B$ and the ghost $c$ to be hermitian; then the antighost $b$
must be antihermitian in order to make ${\cal L}$ hermitian. The rigid
BRST symmetry of ${\cal L}$ under the transformations
$$
\delta A_{\mu}^a=(D_{\mu}c)^a \Lambda ;~~~~~~\delta b^a=\Lambda B^a ;~~~~~~
\delta c^a=\half gf^a_{bc}c^bc^c\Lambda ;~~~~~~\delta B^a=0
\eqno(3.2)
$$  leads to the BRST charge as the space integral of
the time component of the corresponding Noether current. Adding $-c^a$
times the field equation of $A_{\mu}^a$ to the Noether current, the BRST
charge becomes
$$
Q=\int d^3x \left [B_a c^a-\half b_agf^a_{bc}c^bc^c\right ]
\eqno(3.3a)
$$
There is also a conserved ghost charge
$$
Q_{gh}=\int d^3x\left [b_ac^a\right ]
\eqno(3.4)
$$
Redefining the auxiliary field $B_a$ by
$$
d_a=B_a+b_{b}gf^{b}_{ac}c^c
\eqno(3.5)
$$
the fields $d_a$, $b_a$ and $c^a$ all satisfy free-field equations
$$
n.\pd d_a=n.\pd b_a=n.\pd c^a=0
\eqno(3.6)
$$
The BRST charge becomes
$$
Q=\int d^3x \left [d_a c^a+\half b_agf^a_{bc}c^bc^c\right ]
\eqno(3.3b)
$$
and in this form $Q$ and $Q_{gh}$ are manifestly conserved. The fields
$d_a$, $b_a$ and $c^a$ can be expanded as
$$
d_a(t,\x )=\int{{d^3k}\over{(2\pi )^3}}\theta (k_3)
d_a(\k )e^{i(\lambda k_3t+\k .\x )}~~~+\hbox{h c}
$$$$
b_a(t,\x )=\int{{d^3k}\over{(2\pi )^3}}\theta (k_3)
b_a(\k )e^{i(\lambda k_3t+\k .\x )}~~~-\hbox{h c}
$$$$
c^a(t,\x )=\int{{d^3k}\over{(2\pi )^3}}\theta (k_3)
c^a(\k )e^{i(\lambda k_3t+\k .\x )}~~~+\hbox{h c}
\eqno(3.7)
$$
Using Dirac brackets\defref\dirac{
P A M Dirac, {\sl Lectures on quantum mechanics} (Yeshiva University, New York,
1964)
}, the canonical equal-time commutation relations yield
\goodbreak
$$
\{ c^a(\k ),b_b^{\dag}(\k ')\} =-i\delta ^a_b(2\pi )^3\delta (\k -\k ')
$$$$
\{ c^a(\k ),c^{b\dag}(\k ')\}=0=\{ b_a(\k ),b_b^{\dag}(\k ')\}
\eqno(3.8)
$$

The field equation for $d_a$ is just our gauge condition $A_0+\lambda A_3=0$,
and the field equation for $A^a_0$  reads $d_a=(D_jF_{j0})_a$. Further,
because the ghosts decouple, we may work in a subspace of the Fock space
where the kets contain only the ghost vacuum:
$$
|~~~\rangle=|\hbox{nonghost}\rangle \otimes |0~\hbox{ghost}\rangle
{}~~~~\hbox{where}~~~b_a(\k )|0~\hbox{ghost}\rangle =
c^a(\k )|0~\hbox{ghost}\rangle =0
\eqno(3.9)
$$
We shall later consider other possible ghost vacua.
In this subspace we may omit the last term in $Q$, because when we express
it in terms of creation and annihilation operators each term contains
at least one ghost or antighost annihilation operator.
Hence
$$
Q=\int d^3x c^a(\x )(D_iF_{i0})_a=-\int d^3x c^a(\x )(D_i\pi _i)_a
\eqno(3.3c)
$$
Notice that the missing field equation (the Gauss law) is
$(D_iF_{i0})^a=0$, so if this equation were satisfied $Q$ would vanish.
In fact, the Gauss law will be satisfied only as a weak condition,
as we now discuss.

As has been first proposed by Kugo and Ojima\defref\three{
T Kugo and I Ojima, Physics Letters 73B (1978) 459
} we require that physical states be annihilated by $Q$. This generalises the
Gupta-Bleuler condition of QED to general gauges and to nonabelian fields.
As in the previous section, we now pass from the Heisenberg field to the
asymptotic $in$ or $out$ field. We shall omit the colour index on
the field. In the BRST operator only the terms
quadratic in the fields remain, and the terms involving $g$
disappear\ref{\three}. So now, using (2.5), up to an overall
renormalisation\ref{\three}
$$
Q=\int{{d^3k}\over{(2\pi )^3}}\theta (k_3)c^{\dag}(\k )
\; (\o ^2-\lambda ^2k_3^2)\;
p(\k )~~~+\hbox{h c}
\eqno(3.3d)
$$
The terms involving $a_r$ cancel.
When we apply $Q$ to a ket in our subspace (3.9), the term with $c(\k )$
in (3.3d) vanishes.
The kets that are annihilated by $Q$
are those that are annihilated by $p(\k )$,  for all $\k$ (except
when $\k ^2=\lambda ^2k_3^2$). It follows that the Gauss law holds in the
weak sense:
$\langle\hbox{A}|\pd _iF_{i0}|\hbox{B}\rangle =0$ whenever
the kets A and B are annihilated by $Q$.

In general it may be shown\ref{\three} that the solutions to $Q|~~\rangle=0$
have either positive norm, or zero norm. The physical states have positive
norm, whereas the zero-norm states have vanishing inner product with
the physical states and with each other. There may be other conditions
required for a state to be physical, for example in QCD it must have
zero colour. Further, given any ket that represents
a physical state, there are an infinite number of other kets, differing
from it by a piece that is annihilated by $Q$ and has zero norm, all of which
represent the same
physical state. In our case, an example of a zero-norm
state that satisfies $Q|~~\rangle=0$ is $p^{\dag}|0\rangle$, while
states created by the $a_r^{\dag}$ have positive norm. The standard
representative kets for the physical states are obtained by applying
a product of $a_r^{\dag}$ to the vacuum, and adding further pieces
where one or more $p^{\dag}$ is applied to such kets represents the same
physics. The proof that the norm in the subspace $Q|~~\rangle =0$ is
semi-positive-definite relies on the {\it quartet mechanism}
of Kugo and Ojima\ref{\three}.
A quartet consists of two BRST doublets with opposite ghost number. In our
case, the quartet modes are given by $c(\k ),b(\k ),p(\k )$ and $q(\k )$.
Indeed, under BRST transformations of the $in$ or $out$ field
$\delta q(\k )\sim c(\k )$ and $\delta c(\k )=0$ since
$\delta A_{j}=\pd _{j}c$, and $\delta b(\k )\sim B(\k )=d(\k )\sim p(\k )$
and $\delta p(\k )=0$,
while $c(\k )$ and $b(\k )$ have opposite ghost number. All kets in the
asymptotic-field Fock space satisfying $Q|~~\rangle=0$ and $Q_{gh}|~~\rangle
=0$
consist then of the set with the ghost vacuum (which we have been discussing
so far), and further zero-norm states with ghost number zero
constructed from the quartet modes.
An example is the ket
$\left (c^{\dag}b^{\dag}-i(\o ^2-\lambda ^2k_3^2)q^{\dag}p^{\dag}\right )
|\hbox{nonghost}\rangle \otimes |0~\hbox{ghost}\rangle$.

\bigskip
\goodbreak
{\bf 4 Discussion}

1. We require that the Fock vacuum $|0\rangle$ is a physical state. One
allowed vacuum was defined in (2.7) and (3.9). An obvious
alternative\defref\pvl{
P V Landshoff, Physics Letters B227 (1989) 427
},
which is also annihilated by $Q$, is to replace $p,q,b$ and $c$ in
the definition with their hermitian conjugates. This has the effect
of changing $i\epsilon /k_3$ in the propagator (2.10) to
$-i\epsilon /k_3$. If $\lambda <0$ the first choice is more convenient,
and if $\lambda >0$ the second, because then Wick
rotation is possible.

2. We require the vacuum also to have unit norm. We have seen in
Section 3 that if we
define the ghost vacuum to be annihilated by $b$ and $c$, the nonghost vacuum
must be annihilated by $p$. It is
not consistent to require that it be annihilated also by
$p^{\dag}(\k )$, since this would conflict with the vacuum expectation
value of $[p,q^{\dag}]$.
It is not possible to impose the condition that the whole of the Gauss-law
operator $D_i F_{i0}$ annihilates the vacuum (or, indeed, other physical
states) because its equal-time commutator with $A$ is a $c$-number.

3. Physical states should have positive energy. The free-field Hamiltonian
$H_0$,
which governs the time variation of the asymptotic fields according to
$\dot \Phi=i[H_0,\Phi]$, consists of a nonghost part $H_{0a}+H_{0pq}$
and a ghost part. The former is given by
$$
H_{0a}=\int {{d^3k}\over{(2\pi )^3}}\left \{
\half a^{\dag}_r(\k )a_r(\k )+ {{\half (1-\lambda ^2)k_rk_s}\over
{(\lambda\o +k_3)^2}}a^{\dag}_r(\k )a_s(\k )
\right \}
\eqno(4.1a)
$$
and
$$
H_{0pq}=\int {{d^3k}\over{(2\pi )^3}}
\theta (k_3)\left\{ (1-\lambda ^2)(\o ^2-3\lambda ^2k_3^2)p^{\dag}(\k )p(\k )
-\lambda k_3(\o ^2-\lambda ^2k_3^2)(q^{\dag}(\k )p(\k )+p^{\dag}(\k )q(\k ))
\right \}
\eqno(4.1b)
$$
while the ghost part is
$$
H_{0bc}=\int {{d^3k}\over{(2\pi )^3}}
\theta (k_3)i\lambda k_3\left\{ b^{\dag}(\k )c(\k )- c^{\dag}(\k )b(\k )
\right\}
\eqno(4.1c)
$$
As a check, one may verify that $Q$ in (3.3d) commutes with $H_0$.
If we define the vacuum $|0\rangle$ to be annihilated by $Q$, so that
$p(\k )|0\rangle=0$, and also to be an eigenket of $H_0$ with eigenvalue 0,
then we need also $q(\k )|0\rangle=0$ as we have required in (2.7). As we
have already argued, the ket $|0\rangle +p^{\dag}(\k )|\hbox{physical}\rangle$
represents the same physics as $|0\rangle$.  However, it is not an eigenstate
of $H_0$, although it gives the  same expectation value for $H_0$ as
does $|0\rangle$. The same remarks apply to all physical states. The standard
representative kets for the physical states are
annihilated by
$q(\k )$ and are eigenvectors of $H_0$. One may verify that their eigenvalues
are all positive, whatever the value of $\lambda$. For example, the state
$a^{\dag}_r(\k )|0\rangle$ has energy $\o$.

4. If we use the expansion (2.4) of $A_i$, together with $A_0=-\lambda A_3$,
we see that the terms involving $q(\k )$ may be removed by a gauge
transformation $A_{\mu}\to A_{\mu}+\pd _{\mu}\Omega$ with
$(\pd _t+\lambda \pd _3)\Omega =0$. Nevertheless we must retain these
$q(\k )$ modes in the formalism, just as one must keep the longitudinal
polarisations of QED in the Feynman gauge.

5. In our unified treatment, we have seen no basic difference between
the cases where $n^2$ is positive, zero, or negative. For example,
in all cases one can perform Wick rotation. Note, however, that our
treatment does not apply to the gauge $A_3=0$, though an alternative
approach exists for this\defref\spacelike{
P V Landshoff and J C Taylor, Physics Letters B231 (1989) 129
}.

6. It would be interesting to construct the Poincar\'e generators and
investigate how Lorentz boosts relate the
results for different $\lambda$. At first sight it is not clear how
propagators with $i\epsilon/k'_3$ are related to our propagators with
$i\epsilon/k_3$. Perhaps such a relation could extend our analysis to the case
$A_3=0$.

7. In the case of the temporal gauge $A_0=0$ our results agree with previous
work\ref{\pvl}\defref\lazz{
I Lazzizzera, Physics Letters B210 (1988) 188
}. But for the light-cone gauge, $\lambda =\pm 1$, we have not retrieved the
Leibbrandt-Mandelstam prescription\defref\leibbrandt{
S Mandelstam, Nuclear Physics B231 (1983) 149;
G A Leibbrandt, Rev Mod Phys 59 (1987) 1067
} for the propagator: we have
$i\epsilon/k_3$ rather than $i\epsilon/(k_3-k_0)$.
Bassetto et al\defref\five{
A Bassetto, M Dalbosco, I Lazzizzera and R  Soldati, Physical Review
D31 (1985) 2012
} have given a derivation of the light-cone gauge propagator which is very
similar to ours (when $\lambda ^2=1$); at a certain point, however, we each
have to make an assumption that certain poles whose residues are of order
$\epsilon$ may be omitted, and we choose different ones. In our analysis, when
we pass from (2.8) to (2.10)  we have dropped some terms which,
superficially, are of order $\epsilon$. Bassetto et al replace
$$
{1\over{k_0-k_3+i\epsilon\hbox{sign} k_3}}\longrightarrow
{{k_0+k_3}\over{k_0^2-k_3^2+i\epsilon}}
\eqno(4.2)
$$
Such replacements are usually valid when a Wick rotation can be used,
but in a delicate calculation such as that of the Wilson loop
by Andrasi and Taylor\defref\four{
A Andrasi and J C Taylor, Nuclear Physics B375 (1992) 341
}
the difference may be important. It is this subtlety\ref{\two} which has so far
made it impossible to produce a reliable derivation of perturbation
theory in $n.A=0$ gauges.
\bigskip
{\sl We thank T Kugo for helpful correspondence}
\vskip 1truein
\medskip\immediate\closeout\rfile\writestoppt
\baselineskip=14pt{{\bf References}}\bigskip{\frenchspacing%
\parindent=20pt\escapechar=` \input refs.tmp\bigskip}\nonfrenchspacing
\bye